# Hybrid aqueous/ionic liquid electrolyte for high cycle stability and low temperature adaptability lithium-ion battery

Yuewang Yang, Sijing Liu, Zhaowen, Bai and Baoling Huang

Abstract:

Aqueous rechargeable batteries are promising energy storage devices for the high safety, environmental friendliness, and easy assembly. However, their cycle stability and low temperature performance are limited by the narrow electrochemical stability window and the high freezing point of the aqueous electrolytes. Here, a hybrid electrolyte with a wide electrochemical window (2.15V) and a low freezing point (-60 °C) is developed by using EMIMDep as a novel additive. The hydrophobic EMIM+ accumulates on the negative charged electrode and repels the water molecules, thus suppressing the water splitting. Meanwhile, the hydrophilic Dep- forms strong hydrogen bonds with water, thereby reducing the freezing point of the electrolyte. In addition, the hybrid 1 M $LiNO_3$ in $EMIMDep_{20}$-$H_2O_{80}$ electrolytes exhibit high safety and high stability due to the non-flammability, non-volatility, and low toxicity of the EMIMDep compared with other organic additives. Owing to the advantages of the aqueous/EMIMDep electrolyte, the full battery with $LiTi_2(PO_4)_3$ anode and $LiMn_2O_4$ cathode delivers an average voltage of 1.6 V and a specific capacity of 120 mAh/g with a capacity retention of 80% after 500 cycles at 1C. In addition, the full battery working at -35 °C delivers 60% specific capacity of that at room temperature.

**Introduction:**

Aqueous lithium-ion rechargeable batteries (ALIB) are a promising candidate for the next generation of energy storage due to their advantages of high safety, nontoxicity, environmental friendliness and easy assemble[1–4]. However, aqueous lithium-ion batteries suffer from several drawbacks, including low lifespan and limited minimum working temperature, due to the narrow electrochemical window and the high freezing point of the traditional aqueous electrolyte ($LiNO_3$, $Li_2SO_4$ in water)[5–7], which greatly limits the further development of the ALIB. Therefore, simultaneously improving the cycle stability and low temperature performance is critical for the practical application of the ALIB.

Recently, highly concentrated "water-in-salt" (WIS) electrolytes (21m lithium bis(trifluoromethanesulfony) imide in water) expanded the electrochemical window up to 3 V and enabled a 2.3 V lithium-ion battery exhibit high cycle stability[7]. This concept had led to extensive research on WIS electrolytes and a lot of WIS electrolytes have been proposed[8–12]. In the WIS electrolytes, all the water molecules were trapped by the ions and free water molecular does not exist anymore[7]. However, WIS electrolytes show (partially) crystallization precipitation of the salt when they are stored at the room temperature and high freezing point due to the high salt concentration[13], which limits its practical application. Some researchers solve the crystallization problem through tuning the anion of the salts or adding ionic liquids[13,14]. Meanwhile, researchers found that the use of some hydrophobic additives can expand the electrochemical window of the aqueous electrolytes[15–17]. The hydrophobic additives form a protective interphase through absorption and repels the water molecules, thus avoiding the decomposition of the water. For example, adding disodium propane-1,3-disulfonate (PDSS)[16] or sodium dodecyl sulfate (SDS)[17] in aqueous electrolyte could enhance cathodic and anodic limit of the electrolytes. However, these electrolytes still overlooked performance at low temperatures. On the other hand, traditional methods to enable the ALIB to work at low temperature is to add organic antifreezes (such as DMSO, EG, and ACN), which are flammable, volatile and toxic[18–20]. Therefore, it is necessary to develop an aqueous electrolyte that guarantees high cycle stability and low temperature adaptability of the aqueous battery.

Herein, we proposed a hybrid aqueous/ionic liquid electrolyte with a stable electrolyte window from 2.35 V to 4.5 V vs Li/Li+, which covers the working potential of the superior LTP anode (2.4 V vs Li/Li+)[21] and LMO cathode (4.25 V vs Li/Li+)[22] in aqueous battery. The hydrophobic cation, 1-ethyl-3-methylimidazolium+ (EMIM+), in the ionic liquid accumulates on the negatively charged anode and repels the water molecules, thus increasing the cathodic limit[14,23,24]. On the other hand, the hydrophilic anion, diethylphosphonate- (Dep-), in the ionic liquid forms strong hydrogen bonds with water molecules and break the hydrogen bonds among the water molecules, which greatly reduces the freezing point of the electrolytes[25]. In addition, the EMIMDep additive is non-flammable, non-volatile and low in toxicity, making the final electrolyte safer and more reliable. Then, chaotropic salt[26], $LiNO_3$, was dissolve in aqueous/EMIMDep mixture to form the final electrolyte, and we also found that kosmotropic salts (such as $Li_2SO_4$ and LiTFSI)[27] cannot be used due to the low solubility in the mixture. After comprehensive consideration of the effect of EMIMDep content on the electrochemical window and freezing point of the electrolyte, we finally optimized the electrolyte composition to be 20% EMIMDep and 80% water in molar ratio ($EMIMDep_{20}$-$H_2O_{80}$). The full battery constructed with LTP anode and LMO cathode exhibits an average voltage of 1.6 V and retains the 80% initial specific capacity (1C) after 500 cycles at room temperature. In addition, the battery maintains the 60% specific capacity (0.2C) of its room temperature capacity at -35 ºC. A flexible pouch cell battery with high safety and stability was further developed using the LTP anode, LMO cathode, and 1 M $LiNO_3$ in $EMIMDep_{20}$-$water_{80}$ electrolyte.

**Results:**

The freezing points of the aqueous/ionic liquid mixtures are usually higher than -20 °C and it depends more on anion of the ionic liquid[25,28]. However, some special ionic liquids (such as EMIMDep and EMIMAc) can form mixtures with water, which achieve an ultra-low freezing point. These anions of these ionic liquid form strong hydrogen bonds with water molecules and weak the hydrogen bonds among the water molecules, and thus significantly decrease the freezing temperature of the mixtures. Among these ionic liquids, EMIMDep was chosen for the high stability and proper PH (=7) as the additives to form electrolytes with water and $LiNO_3$. The Fig. 1A displays the structure of the 1 M $LiNO_3$ in hybrid aqueous/EMIMDep electrolytes, and the P=O in the anion interact strongly with the water molecules. The freezing points of different compositions are determined by differential scanning calorimetry (DSC), as shown in Fig. 1B. The exothermic peaks in the cooling scan for all the samples corresponding to the freezing points (crystallization). Combined with previous freezing point results in Liu's work[25], a low freezing point of -60 °C is found at around 4:1 molar ratio between EMIMDep and water. Meanwhile, the freezing point of the traditional aqueous electrolytes (1 M $LiNO_3$ or $Li_2SO_4$ in water) is around -20 °C, which limits the battery performance at temperature lower than -20 °C. On the other hand, the increase of the EMIMDep content after 20% in molar ratio decreases the ionic conductivity and 1 M $LiNO_3$ cannot dissolve in the mixtures due to the limit solubility when the EMIMDep content is high. Therefore, we start from the 1 M $LiNO_3$ in $EMIMDep_{20}$-$H_2O_{80}$ electrolyte to demonstrate our battery, which meets the need for low freezing point and sufficient lithium ions. In addition, kosmotropic salts (such as 1 M $Li_2SO_4$ and LiTFSI) were found to cannot dissolve in the mixture (Fig. S1). The water molecules in the aqueous/EMIMDep form a lot of hydrogen bonds with EMIMDep, and there are not enough excess hydrogen bonds to dissolve the kosmotropic salts, which requires more hydrogen bonding when dissolving compared with chaotropic salts. The variation of the ionic conductivity with temperatures for different electrolytes is shown in Fig. 1C and the corresponding EIS results are shown in (Fig. S2, 3, and 4). The ionic conductivity of the 1M $LiNO_3$ in $EMIMDep_{20}$-$H_2O_{80}$ is 5.6 mS/cm at room temperature and maintains 0.015 mS/cm at -40 °C. Meanwhile, the ionic conductivity of the $EMIMDep_{20}$-$H_2O_{80}$ is always higher than that corresponding one with 1 M $LiNO_3$, this is because the addition of $LiNO_3$ increases the viscosity of the electrolytes and thus reduces the ionic conductivity. On the contrary, 1 M $LiNO_3$ in water electrolyte exhibits the highest ionic conductivity at the temperatures above the -20 ºC, but the ionic conductivity cannot maintain at the temperature below -20 ºC due to the freezing of the electrolytes.

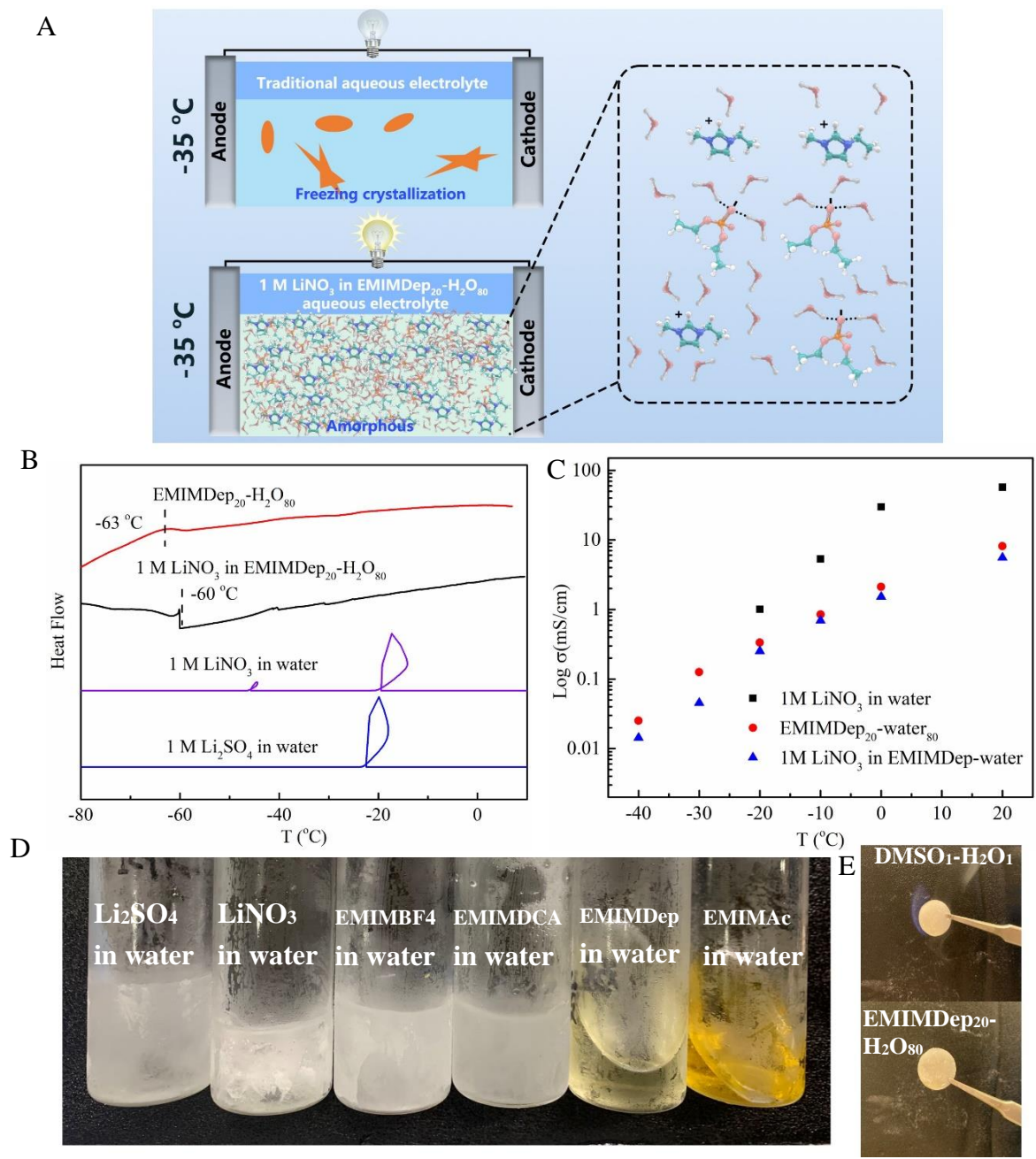

**Fig. 1.** (A) Structure and the interaction of the 1 M LiNO$_3$ in EMIMDep$_{20}$-H$_2$O$_{80}$ electrolyte. (B) Freezing points of the different electrolytes. (C) Conductivity of different electrolytes at different temperatures. (D) The state of different electrolytes at -40 ºC. (E) Flammability of the DMSO$_1$-H$_2$O$_1$ and EMIMDep$_{20}$-H$_2$O$_{80}$.

The photograph of the different electrolytes at -40 °C is presented in the Fig. 1D. Obviously, the traditional aqueous electrolytes (1 M $Li_2SO_4$ and $LiNO_3$ in water) and hybrid aqueous/ ionic liquid mixtures with conventional ionic liquid ($EMIMBF_4$ and EMIMDCA in water) freeze at -40 °C. On the contrary, the hybrid aqueous/ionic liquid mixtures with special ionic liquid (EMIMDep and EMIMAc in water) maintain the liquid state at this temperature, which is beneficial for batteries to operate at low temperature.

Traditional low temperature aqueous battery achieved through adding organic antifreeze additives, for example, DMSO, ACN and EG, and the electrolytes become flammable when the organic additives exceed a certain level. However, the EMIMDep additive is non-flammable and stable, enabling the high safety of the battery. Therefore, we also conducted the flammability test of the electrolyte. The $EMIMDep_{20}$-$water_{80}$ is not flammable when exposed to fire, in contrast, the low temperature electrolyte with DMSO additives is ignited (Fig. 1E). Thus, the battery using 1 M $LiNO_3$ aqueous/EMIMDep exhibits improved safety compared with traditional low temperature aqueous battery.

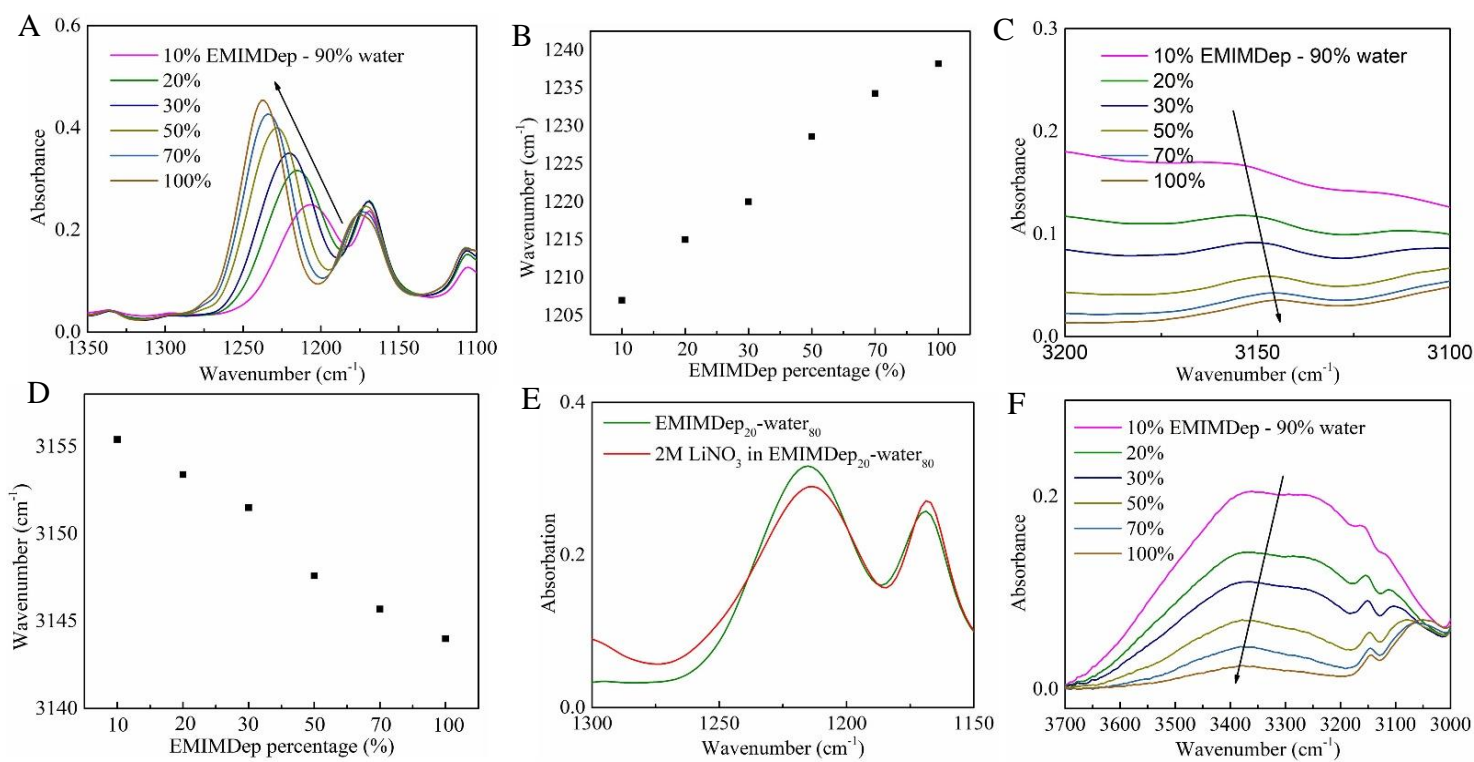

**Fig. 2.** FTIR spectra of the EMIMDep/$H_2O$ mixtures with different molar fractions of the EMIMDep (A) P=O band in Dep- of the EMIMDep. (B) The shift of the P=O band. (C) $C_2$-H band in EMIM+ of the EMIMDep. (D) The shift of the $C_2$-H band. (E) P=O band change with and without LiNO$_3$ additives. (F) O-H stretching bands of H$_2$O at different molar fractions of the EMIMDep in the mixture.

The Fourier-transform infrared spectroscopy (FTIR) results of the hybrid aqueous/EMIMDep presented in Fig. 2 show the stretching vibration for P=O in Dep-, C$_2$-H in EMIM+, and O-H in water with different molar fractions of the EMIMDep. The full FTIR spectra results are presented in Fig. S5. In Fig. 2A, the band at 1207 cm$^{-1}$ is assigned to asymmetric stretching vibration of P=O[29,30]. As the content of the EMIMDep increases, this band shifts to higher wavenumbers (blue shift), which corresponding to the hydrogen bonds between the H in the water molecular and O in the P=O in the Dep- (P=O···H-O). There is a large degree of shift in the band of P=O, from 1207 to 1240 cm$^{-1}$ as the EMIMDep content increases from 10% to 100% in molar fraction, which means that this interaction (P=O···H-O) is very strong. Meanwhile, the water molecules also form hydrogen bonds with EMIM+ cation. The band at 3155 cm$^{-1}$ is assigned to C$_2$-H stretch[31] in EMIM+ and the H in the C$_2$-H can form hydrogen bonds with the O in the water molecules (C$_2$-H···O-H). This band shifts to lower wavenumber (red shift) as the content of the EMIMDep increases, and the shift from 3155 to 3144 cm$^{-1}$ as the EMIMDep content increases from 10% to 100% in molar fraction. The shift of C$_2$-H is much smaller than that of P=O as the EMIMDep content increases, which means that water molecules are more inclined to form hydrogen bonds with anions. This also corresponds to the anion in the ionic liquid that is the main factor in lowering the freezing point. On the other hand, Fig. 2E shows the slightly change of the band at 1217 cm$^{-1}$ (P=O in Dep-) when the LiNO$_3$ added to the mixture. This means that adding LiNO3 does not change the structure of the mixture, which enables the final electrolyte maintains the low freezing point. Fig. 2F shows the band of O-H stretching vibration shifts to higher wavenumber as the EMIMDep content increases. This shift indicates that hydrogen bonding among the water molecules have been weaken, which benefits to preventing forming crystals for water molecules and achieving the low freezing point of the electrolytes[18].

To further demonstrate the low temperature performance of the 1 M LiNO$_3$ in EMIMDep$_{20}$-water$_{80}$ electrolyte in practical battery applications, we choose the proper electrodes (LTP anode and LMO cathode), whose superior properties have been discussed in the previous article[21,22]. The battery using 1 M LiNO$_3$ in EMIMDep$_{20}$-water$_{80}$ electrolyte provides a specific capacity of around 60 mAh/g at the current of 0.2C at -35 °C, and there is almost no decay after

50 cycles (Fig. 3A). Fig. 3B shows the charge-discharge profile of the battery at -35 °C and room temperature, in which the specific capacity of the battery at -35 °C is 58% of that at room temperature (Fig. 3B). The EIS results for the batteries using 1 M LiNO$_3$ in EMIMDep$_{15}$-water$_{85}$, EMIMDep$_{20}$-water$_{80}$ and EMIMDep$_{25}$-water$_{75}$ at -35 °C are shown in the Fig. S6. The resistance of the batteries increases as the EMIMDep content increases due to the lower conductivity when the EMIMDep content is high. On the other hand, the battery can still power the LED at -35 °C, as shown in the Fig. 3C.

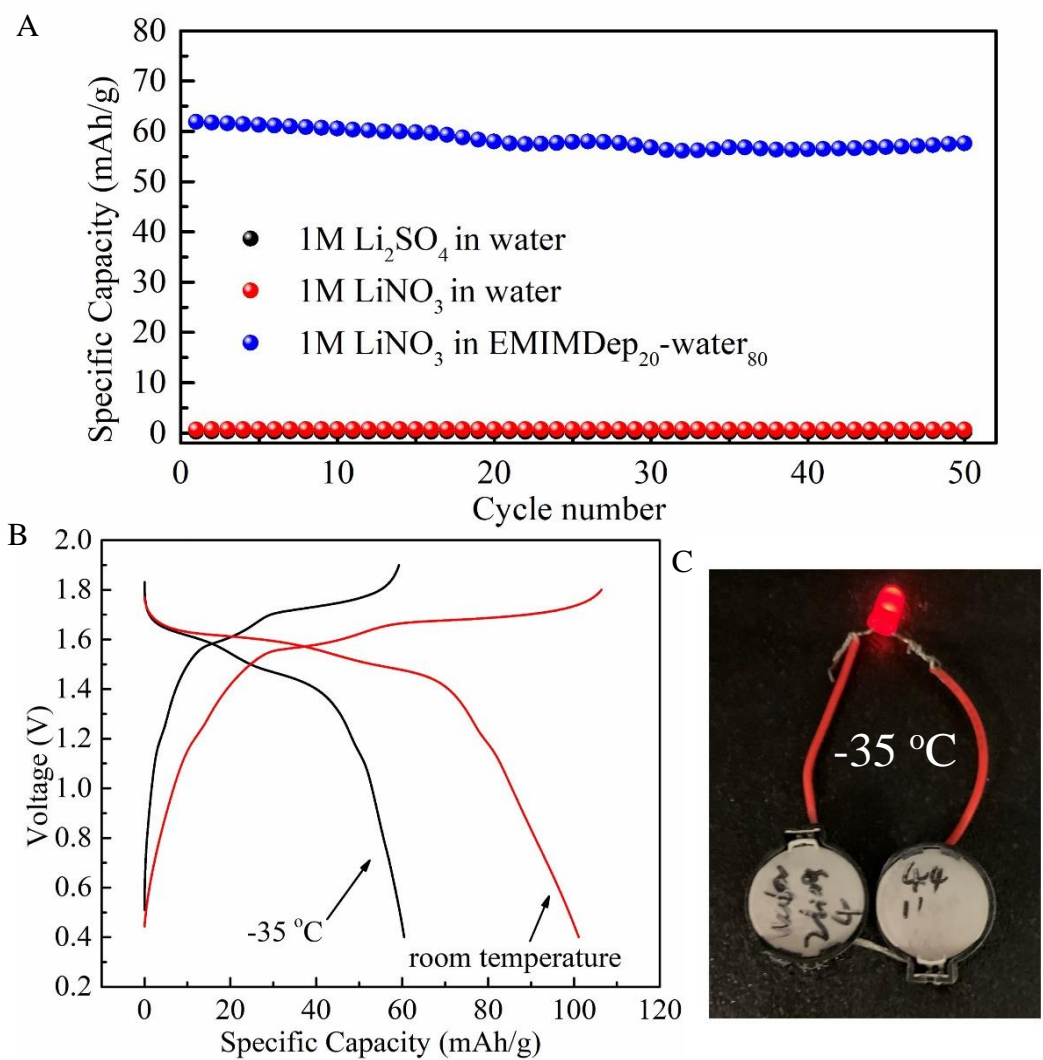

**Fig. 3.** (A) Cycle performance of the LMO/LTP battery using 1 M Li$_2$SO$_4$ in water, 1 M LiNO$_3$ in water, and 1 M LiNO$_3$ in EMIMDep$_{20}$-H$_2$O$_{80}$ electrolytes at -35 °C. (B) Charge-discharge profiles of the battery using 1 M LiNO$_3$ in EMIMDep$_{20}$-H$_2$O$_{80}$ at the current of 0.2C at -35 °C and room temperature. (C) Photograph of the battery powering a LED at -35 °C.

The electrochemical properties are investigated through linear sweep voltammetry (LSV) test to further understand the cycle stability of the 1 M LiNO$_3$ in EMIMDep$_{20}$-H$_2$O$_{80}$ electrolyte in LMO/LTP system. The LSV test was conducted with stainless steel mesh as the counter and working electrodes, and Ag/AgCl in 3 M KCl aqueous electrolyte as the reference electrode at the scan rate of 1 mV/s. The stainless steel is chosen because the current collector we used is stainless steel, therefore, the electrochemical window of the electrolyte is measured based on that. In addition, the potential vs Ag/AgCl has been converted to vs Li/Li+ for convenience. Fig. 4A shows the cathodic and anodic scans of the different electrolytes. The anodic limits of the three electrolytes are all around 4.5 V vs Li/Li+, which are higher than the working voltage (4.25 V vs Li/Li+) of the LMO cathode. This means that all the three electrolytes are electrochemical stable for cathode side. On the other hand, the cathodic limits of the three electrolytes are 2.35, 2.9 and 2.7 V vs Li/Li+ for LiNO$_3$ in EMIMDep$_{20}$-H$_2$O$_{80}$, LiNO$_3$ in water and Li$_2$SO$_4$ in water, respectively. The cathodic limits of the LiNO$_3$ in water (2.9 V vs Li/Li+) and Li$_2$SO$_4$ in water (2.7 V vs Li/Li+) are lower than the working potential of the LTP anode (2.4 V vs Li/Li+), which means these two electrolytes are not stable on the LTP anode and hydrogen evolution happens prior the insertion of the Li+ to LTP. In the contrast, the cathodic limits of the LiNO$_3$ in EMIMDep$_{20}$-H$_2$O$_{80}$ is lower than 2.4 V, which guarantees the electrochemical stability of the electrolyte in the battery. The expanded cathodic electrochemical window comes from the two factors: 1, the water molecules forms hydrogen bonds with added EMIMDep, and the proportion of the free water molecules is reduced. 2, the hydrophobic EMIM+, which is conformed in previous FTIR results, accumulated on the negative charged anode, and repels the water molecules (Fig. 4B). The effect of EMIMDep content and lithium salt to the electrochemical window are presented in Fig. 4C, and D. There is almost no change for anodic limits as the EMIMDep content increase, this is because that the water molecules move together with the hydrophilic Dep- (strong hydrogen bonds) and decompose at the positive charged cathode (Fig. 4B). On the contrary, the cathodic limits change a lot, from 2.6 V to 2.16 V vs Li/Li+ when the EMIMDep content increases from 10% to 30% in molar fraction. The accumulated EMIM+ repels the water molecules and this effect increases as the EMIM+ content increases. The interaction between the water molecules and Li+, NO$_3^-$ is also very strong and the water molecules move together with these ions, therefore, the effect of the adding of LiNO$_3$ to both anodic and cathodic limits is small (Fig. 4D).

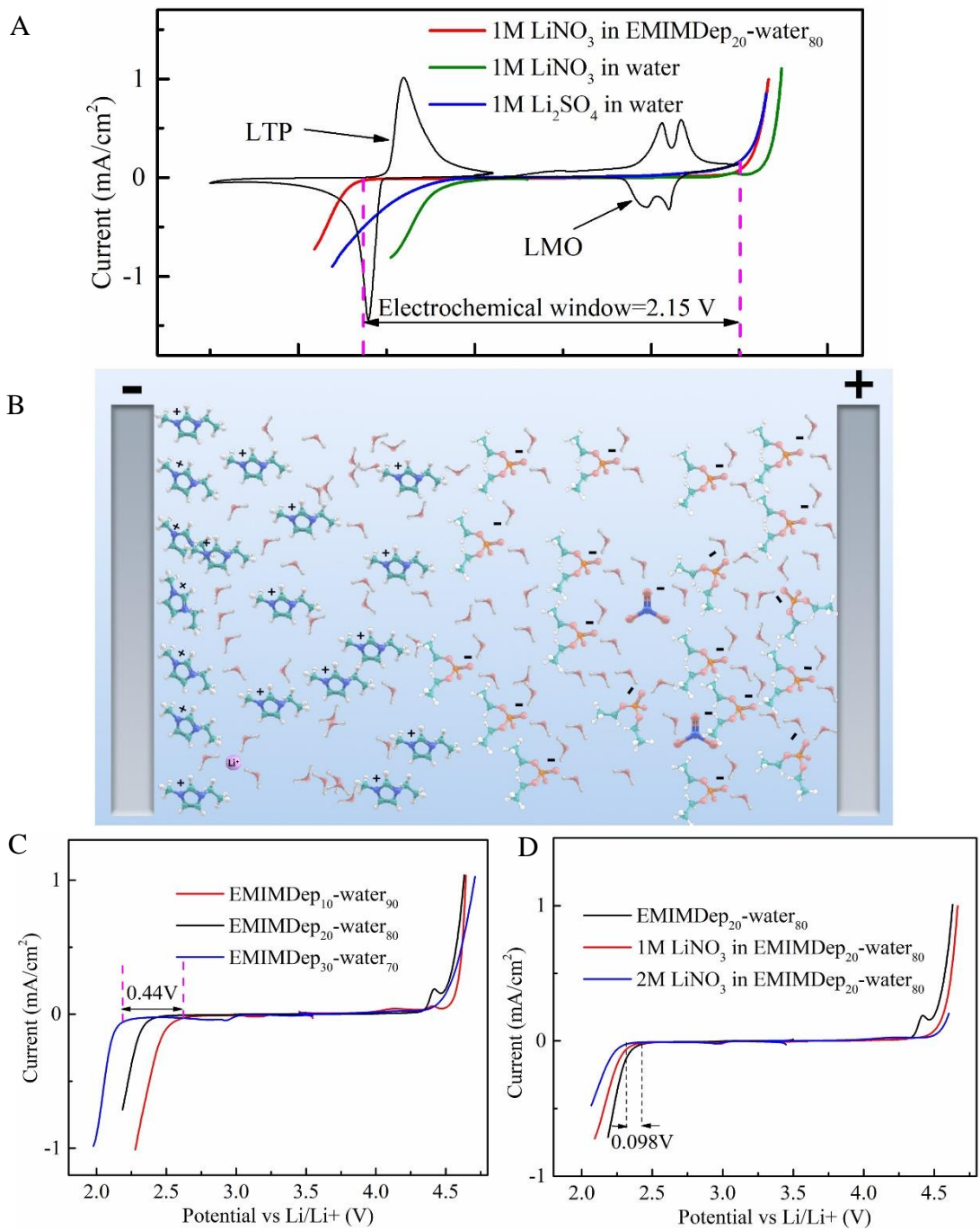

**Fig. 4.** (A) The electrochemical window of the 1 M LiNO$_3$ in EMIMDep$_{20}$-H$_2$O$_{80}$, 1 M LiNO$_3$ in water, and 1 M Li$_2$SO$_4$ in water electrolytes. (B) The mechanism of the increased cathodic limit of the 1 M LiNO$_3$ in EMIMDep$_{20}$-H$_2$O$_{80}$, and the state of ions and water molecules in the electrolyte in an electric field. (C) The change of the electrochemical window for the EMIMDep/H$_2$O mixtures with different molar fractions of the EMIMDep. (D) The effect of the additive of the LiNO$_3$ to the electrochemical window of the EMIMDep/H$_2$O mixtures.

The electrochemical stability of the 1 M $LiNO_3$ in $EMIMDep_{20}$-$H_2O_{80}$ were further evaluated in the galvanostatic charge-discharge test of the LMO/LTP battery. The battery using 1 M $LiNO_3$ in $EMIMDep_{20}$-$H_2O_{80}$ maintains 80% of its initial specific capacity (115 mAh/g) after 500 cycles at the current of 1C at room temperature (Fig. 5A). On the contrary, the batteries using 1 M $LiNO_3$ in water and 1 M $Li_2SO_4$ in water retain only 33% and 44% of their initial specific capacity, respectively. In addition, the coulombic efficiencies for the first cycle of the batteries using 1 M $LiNO_3$ in water, 1 M $Li_2SO_4$ in water, and 1 M $LiNO_3$ in $EMIMDep_{20}$-$H_2O_{80}$ electrolytes are 63%, 72%, and 85%, respectively (Fig. S7). These results confirm the superior electrochemical stability of the 1 M $LiNO_3$ in $EMIMDep_{20}$-$H_2O_{80}$ electrolyte. The low freezing point mixture using EMIMAc were also tested as electrolyte in the battery and the specific capacity decays fast, and we assume that the side reactions are serious between the acetate anion and electrodes (Fig. S8). Meanwhile, a flexible pouch cell was assembled using a stainless-steel mesh current collector to power a LED, and the high safety and low toxicity properties of the electrolytes and electrodes makes it possible to supply power for the human wearable devices (Fig. 5B). The charge-discharge profiles of the LMO vs lithium metal, LTP vs lithium metal, and LMO vs LTP are shown in Fig. 5C, and the synthesized LTP exhibits a stable platform at the voltage of 2.5 V vs Li/Li+. Fig. 5D exhibits the rate performance of the batteries using 1 M $LiNO_3$ in water and 1 M $LiNO_3$ in $EMIMDep_{20}$-$H_2O_{80}$ at room temperature. The specific capacities of the 1 M $LiNO_3$ in $EMIMDep_{20}$-$H_2O_{80}$ electrolyte based full batteries are 113, 103, 85, and 60 mAh/g at the currents of 1C, 2C, 4C and 8C, respectively. These specific capacities are close to that using 1 M $LiNO_3$ in water electrolyte at the current below 4C, which are 116, 112, and 100 mAh/g at the currents of 1C, 2C and 4C, respectively. However, the specific capacity of the battery using 1 M $LiNO_3$ in water electrolyte is 90 mAh/g, which is significantly larger than that using 1 M $LiNO_3$ in $EMIMDep_{20}$-$H_2O_{80}$. This difference comes from the higher conductivity of the 1 M $LiNO_3$ in water. The electrochemical stability of the batteries is analysed through EIS tests of the full batteries after different cycles (Fig. 5E and F), and the EIS test is conducted at a frequency of $10^5$ to 0.1 Hz when the battery is discharged to 0.4 V and stands for an hour to reach a stable open circuit potential. The equivalent circuit models for fitting the EIS curves are presented in the insets of the Fig. 5E and F and the fitting results are summarized in Table S1 and Table S2. The internal resistance ($Rs$) in the equivalent circuit models corresponds to the high frequency intercept with real axis, which represents the ionic resistance of the electrolyte and the intrinsic resistance of the electrodes. The charge transfer resistances for cathode (*Rctc*) and anode (*Rcta*) corresponds to the medium frequency semicircle and low frequency semicircle in the equivalent circuit models,

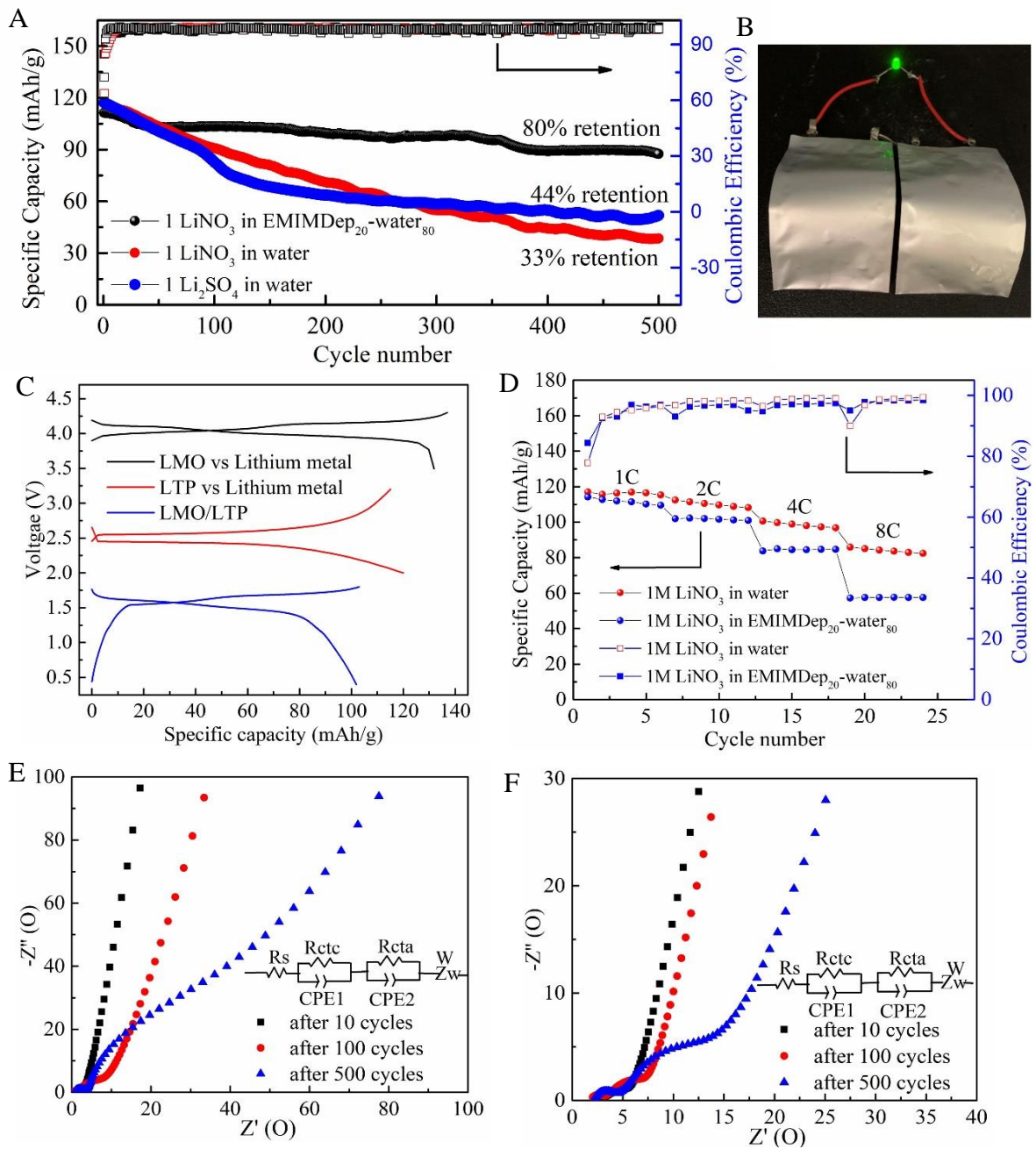

**Fig. 5.** (A) Cycling stability of the batteries using 1 M $LiNO_3$ in $EMIMDep_{20}$-$H_2O_{80}$, 1 M $LiNO_3$ in water, and 1 M $Li_2SO_4$ in water electrolytes. (B) Photograph of the LED powered by the flexible pouch cell battery. (C) Galvanostatic charge-discharge profiles of the LMO vs lithium metal battery, LTP vs lithium metal battery, and LMO/LTP battery. (D) Rate performance of the batteries using 1 M $LiNO_3$ in $EMIMDep_{20}$-$H_2O_{80}$, and 1 M $LiNO_3$ in water electrolytes. (E) EIS results of the LMO/LTP battery using 1 M $LiNO_3$ in water electrolyte

after different cycles. (F) EIS results of the LMO/LTP battery using 1 M $LiNO_3$ in $EMIMDep_{20}$-$H_2O_{80}$ electrolyte after different cycles.

respectively, which are determined by the kinetic resistance at the interface between cathode and electrolytes, anode and electrolytes, respectively. The $R_{cta}$ of the battery using 1 M $LiNO_3$ in water electrolyte are 3.396, 14.23, and 111.6 Ω after 10, 100, and 500 cycles, respectively, and the significant change after 500 cycles represents the instability of the interface between anode and electrolytes. On the contrary, the $R_{cta}$ of the battery using 1 M $LiNO_3$ in $EMIMDep_{20}$-$H_2O_{80}$ electrolyte are 6.056, 8.227, and 20.77 Ω after 10, 100, and 500 cycles, respectively. This slight change of the $R_{cta}$ after 500 cycles proves the high stability of the electrolyte in contact of the LTP anode. Overall, the stability of the battery using 1 M $LiNO_3$ in $EMIMDep_{20}$-$H_2O_{80}$ is conformed in galvanic charge-discharge test and EIS test, pathing the way for the practical application of the aqueous lithium-ion battery.

**Conclusion**

In summary, LMO/C-LTP aqueous batteries using a hybrid aqueous/ionic liquid (1 LiNO$_3$ in EMIMDep$_{20}$-H$_2$O$_{80}$) electrolyte were developed to achieve the high cycle stability, 80% specific capacity retention after 500 cycles at 1 C, and superior low temperature adaptability, 60% capacity retention at -35 °C, simultaneously. A superior electrolyte with wide electrochemical window (2.15 V) and low freezing point (-60 °C) is achieved by appropriate selection of anions and cations of the ionic liquids. The hydrophobic EMIM+ in the EMIMDep accumulates on the negative charged anode and repels the water, thus pushing the cathodic limit to 2.35 vs Li/Li+. On the other hand, the hydrophilic Dep- form strong hydrogen bonds with water molecules and damage the hydrogen bonds network among water molecules, thus lowering the freezing point of the mixture. Additionally, the 1 M LiNO$_3$ in EMIMDep$_{20}$-H$_2$O$_{80}$ exhibits the high safety, low toxicity, and high stability due to the merits, non-flammability, and non-volatility, of the EMIMDep. A flexible pouch cell battery was further demonstrated to explore the possibilities of supplying power for human wearable devices.

**Experimental Section**

**Preparation of aqueous/EMIMDep electrolyte:**

1-ethyl-3-methylimidazolium diethyl phosphate (EMIMDep) and LiNO$_3$ were purchased from Aladdin. The hybrid aqueous/EMIMDep mixture were prepared by mixing EMIMDep and water at a molar ratio of EMIMDep: H$_2$O = 1:9, 2:8, 3:7, 5:5, 7:3, respectively. Then, different concentrations LiNO$_3$ were dissolved in the mixture through sufficiently stirring to form the final electrolytes. Meanwhile, 1 M Li$_2$SO$_4$ and LiNO$_3$ was dissolved in the water through sufficiently stirring to form two electrolytes as the control group.

**Synthesis of C-LTP electrodes:**

Stoichiometric amount of Li$_2$CO$_3$, (Aladdin), TiO$_2$ (Aladdin) and (NH$_4$)$_2$HPO$_4$ (Aladdin) were ball-milled for 3 hours to achieve thorough mixing. The calcined the mixture at 500 °C for 5 hours in air to remove the NH$_4$. Next, 900 °C heat treatment in air was applied for 24 hours to get the final LTP. Then, as-synthesized LTP was ball-milled with glucose in a mass ratio of 9:1 using the water as the solvent. Finally, the glucose was carbonized at 800 °C for 5 hours in flowing Ar. The XRD of the C-LTP is shown in Fig. S9.

**Battery assembly:**

70 wt% active materials (LTP and LMO) with 10 wt% PVDF, and 20 wt% carbon black in NMP solution was mixed to form the slurries of cathode and anode. The slurries were evenly coated on circular iron sheets with a radius of 0.7 cm followed by baking in a vacuum oven at 100 °C for 12 hours to get the final electrodes. The final mass loadings of the cathode and anode active material are 3 mg/cm$^2$ and 10 mg/cm$^2$, respectively. The specific capacity of the battery was calculated based on the cathode active material. Finally, 2025-coin cell batteries were assembled using hybrid aqueous/EMIMDep electrolyte, LMO cathode, C-LTP anode and glass fibre (Aladdin) separator. The flexible pouch cell demonstration was prepared through similar method with coin cell. The difference is that the current collector is replaced by a flexible stainless-steel mesh.

**Characterizations:**

X-ray diffraction (XRD) patterns of C-LTP were recorded on X'pert Pro in 2-theta ranging from 10 to 90 degree. Fourier-transform infrared spectroscopy (FTIR) of the hybrid aqueous/EMIMDep were obtained using a Bruker Vertex 70 Hyperion 1000 spectrometer

(ATR) at the wavenumber range of 4000 to 400 cm$^{-1}$. Differential scanning calorimetry was conducted with a DSC TA Q 1000. The samples were scanned from 20 to -80 °C at a cooling rate of 1 °C/min in N$_2$ atmosphere, then the samples were held at – 80 °C for ten minutes to freeze the samples. Finally, the samples were scanned from -80 to 20 °C at a heating rate of 1 °C/min in N$_2$ atmosphere. The flammability test of the electrolytes was performed by igniting the hybrid aqueous/EMIMDep soaked in glass fibre, and the control group was DMSO$_1$-H$_2$O$_1$ according to the other article[18]. The ionic conductivity of the electrolytes was measured at different temperatures using electrochemical workstation with a DJS-1C probe (nominal constant: 1.0 cm$^{-1}$), and the setup was calibrated with a standard solution (ion conductivity: 12.88 mS/cm at 25 °C). The samples were left standing at each temperature for at least 2 hours before ionic conductivity test.

**Electrochemical Measurements:**

The electrochemical window of the electrolytes was measured through linear sweep voltammetry (LSV) test at a scan rate of 1 mV/s. Here, a three-electrode system was used with stainless steel as the working and counter electrode, and Ag/AgCl (in 3 M KCl aqueous solution) as the reference electrode. Then, the electrochemical window was converted to Li/Li+ reference for convenience. The working potential of the LMO and LTP was measured in a 2025-coin cell with lithium metal as anode and traditional organic electrolyte (1M LiPF$_6$ in EC/EMC) as electrolyte at the voltage range of 3.5 V - 4.3 V, 2.0 V- 3.2 V, respectively. The long cycle performance of the batteries was tested on a battery test system (XINWEI, CT2001A) in a voltage range of 0.4-1.8 V at the current of 1 C. The low temperature performance of the batteries was tested on the same battery test system with the battery exposing to different temperatures at the voltage range of 0.4-1.9 V at the current of 0.2 C. The battery temperature was controlled by a polycarbonate open bath system. The electrochemical impedance spectroscopy (EIS) of the full batteries after different cycles and under different temperatures were performed on a Metrohm Autolab PGSTAT302N electro-chemical station.

1. Hu, L. & Xu, K. Nonflammable electrolyte enhances battery safety. *Proc. Natl. Acad. Sci. U. S. A.* **111**, 3205–3206 (2014).

2. Kim, H. *et al.* Aqueous rechargeable Li and Na ion batteries. *Chem. Rev.* **114**, 11788–11827 (2014).

3. Liu, J., Xu, C., Chen, Z., Ni, S. & Shen, Z. X. Progress in aqueous rechargeable batteries. *Green Energy Environ.* **3**, 20–41 (2018).

4. Head, J. Relative To the Crater Diameter Is Verv Tvoical. *Science (80-. ).* **264**, 1–4 (1994).

5. Ramanujapuram, A. & Yushin, G. Understanding the Exceptional Performance of Lithium-Ion Battery Cathodes in Aqueous Electrolytes at Subzero Temperatures. *Adv. Energy Mater.* **8**, 1–8 (2018).

6. Luo, J. Y., Cui, W. J., He, P. & Xia, Y. Y. Raising the cycling stability of aqueous lithium-ion batteries by eliminating oxygen in the electrolyte. *Nat. Chem.* **2**, 760–765 (2010).

7. Suo, L. *et al.* 'Water-in-salt' electrolyte enables high-voltage aqueous lithium-ion chemistries. *Science (80-. ).* **350**, 938–943 (2015).

8. Suo, L. *et al.* 'water-in-Salt' electrolytes enable green and safe Li-ion batteries for large scale electric energy storage applications. *J. Mater. Chem. A* **4**, 6639–6644 (2016).

9. Dubouis, N. *et al.* The role of the hydrogen evolution reaction in the solid-electrolyte interphase formation mechanism for ": Water-in-Salt " electrolytes. *Energy Environ. Sci.* **11**, 3491–3499 (2018).

10. Suo, L. *et al.* Advanced High-Voltage Aqueous Lithium-Ion Battery Enabled by "Water-in-Bisalt" Electrolyte. *Angew. Chemie* **128**, 7252–7257 (2016).

11. Ko, S. *et al.* Lithium-salt monohydrate melt: A stable electrolyte for aqueous lithium-ion batteries. *Electrochem. commun.* **104**, 106488 (2019).

12. Suo, L. *et al.* "Water-in-Salt" Electrolyte Makes Aqueous Sodium-Ion Battery Safe, Green, and Long-Lasting. *Adv. Energy Mater.* **7**, 1–10 (2017).

13. Reber, D., Kühnel, R. S. & Battaglia, C. Suppressing Crystallization of Water-in-Salt

Electrolytes by Asymmetric Anions Enables Low-Temperature Operation of High-Voltage Aqueous Batteries. *ACS Mater. Lett.* **1**, 44–51 (2019).

14. Dou, Q. *et al.* "Water in salt/ionic liquid" electrolyte for 2.8 V aqueous lithium-ion capacitor. *Sci. Bull.* **65**, 1812–1822 (2020).

15. Stojković, I. B., Cvjetićanin, N. D. & Mentus, S. V. The improvement of the Li-ion insertion behaviour of Li1.05Cr0.10Mn1.85O4 in an aqueous medium upon addition of vinylene carbonate. *Electrochem. commun.* **12**, 371–373 (2010).

16. Miyazaki, K. *et al.* Enhanced resistance to oxidative decomposition of aqueous electrolytes for aqueous lithium-ion batteries. *Chem. Commun.* **52**, 4979–4982 (2016).

17. Hou, Z. *et al.* Surfactant widens the electrochemical window of an aqueous electrolyte for better rechargeable aqueous sodium/zinc battery. *J. Mater. Chem. A* **5**, 730–738 (2017).

18. Nian, Q. *et al.* Aqueous Batteries Operated at −50 °C. *Angew. Chemie - Int. Ed.* **58**, 16994–16999 (2019).

19. Nguyen Thanh Tran, T., Zhao, M., Geng, S. & Ivey, D. G. Ethylene Glycol as an Antifreeze Additive and Corrosion Inhibitor for Aqueous Zinc-Ion Batteries. *Batter. Supercaps* **202100420**, 1–8 (2022).

20. Sun, Y. *et al.* Towards the understanding of acetonitrile suppressing salt precipitation mechanism in a water-in-salt electrolyte for low-temperature supercapacitors. *J. Mater. Chem. A* **8**, 17998–18006 (2020).

21. Xu, T. *et al.* Nanostructured LiTi2(PO4)3 anode with superior lithium and sodium storage capability aqueous electrolytes. *J. Power Sources* **481**, 1–11 (2021).

22. Lengyel, M., Zhang, X. & Axelbaum, R. Electrochemical Performance of LiMn 2 O 4 Powders Prepared with a Novel Honeycomb Burner . *ECS Meet. Abstr.* **MA2011-01**, 204–204 (2011).

23. Feng, G., Jiang, X., Qiao, R. & Kornyshev, A. A. Water in ionic liquids at electrified interfaces: The anatomy of electrosorption. *ACS Nano* **8**, 11685–11694 (2014).

24. Zhang, Y., Ye, R., Henkensmeier, D., Hempelmann, R. & Chen, R. "Water-in-ionic liquid" solutions towards wide electrochemical stability windows for aqueous rechargeable batteries. *Electrochim. Acta* **263**, 47–52 (2018).

25. Liu, Y. *et al.* Freezing Point Determination of Water-Ionic Liquid Mixtures. *J. Chem. Eng. Data* **62**, 2374–2383 (2017).

26. Assaf, K. I. & Nau, W. M. The Chaotropic Effect as an Assembly Motif in Chemistry. *Angew. Chemie - Int. Ed.* **57**, 13968–13981 (2018).

27. Naseem, B., Arif, I. & Jamal, M. A. Kosmotropic and chaotropic behavior of hydrated ions in aqueous solutions in terms of expansibility and compressibility parameters. *Arab. J. Chem.* **14**, 103405 (2021).

28. Osmanbegovic, N., Yuan, L., Lorenz, H. & Louhi-Kultanen, M. Freeze concentration of aqueous [DBNH][OAC] ionic liquid solution. *Crystals* **10**, 1–14 (2020).

29. Liu, H., Chen, S., Li, X., Zhao, R. & Sun, Y. Preparation of [EMIM]DEP/2C3H4O4 DESs and its oxidative desulfurization performance. *Sep. Sci. Technol.* **56**, 558–566 (2021).

30. Shi, L., Chen, Y., Gong, X. & Yu, X. Synthesis and characterization of quaternary ammonium salt tertiary amide type sodium hydroxypropyl phosphate asphalt emulsifier. *Res. Chem. Intermed.* **45**, 5183–5201 (2019).

31. Booth, R. S. *et al.* Identification of multiple conformers of the ionic liquid [emim][tf2n] in the gas phase using IR/UV action spectroscopy. *Phys. Chem. Chem. Phys.* **18**, 17037–17043 (2016).